# CRITICAL PHENOMENA IN THE DYNAMICAL VISIBILITY GRAPH

A.A.Snarskii[1,2], I.V.Bezsudnov[3]

[1] National Technical University "Kiev Polytechnic Institute", Kiev, Ukraine

[2] Institute for information recording, NAS Ukraine, Kiev, Ukraine

[3] JSC Nauka-Service, Moscow, Russia

We have investigated the time series by the mapping them to the complex network. We have studied the behavior of the relative number of clusters in dynamic visibility graphs near the critical value of the angle of view. Time series of different nature both artificial and experimental were numerically investigated. In all cases, the dependence of the relative number of clusters on the proximity to the critical angle of view had a power law behavior. Thus, it was shown that there is a similarity between the behavior of the relative number of clusters and the order parameter in the second-order phase transition theory and percolation theory. Each time series is characterized by its own value of the critical index - an analog of the critical index in second-order phase transition theory.

The investigation of time series with complex or fractal structure draws continuous interest due to the large application potential of such studies. Huge range of phenomena from heart rate to earthquakes events [1-8] can be presented in form of time series. Different methods can be used to characterize such time series: common statistics (mean values, moments of distributions and more), the various spectrum examinations (eg. 1/f power spectrum noise), the determination of the fractal dimension of time series etc. Some methods allow detecting quite nontrivial properties, such as multifractality, the existence of strange attractors.

In this paper we show that the analysis of time series using dynamical visibility graph method [9] extracts the property that behaves similarly to the order parameter in second-order phase transitions theory.

Recently, under active development is a group of methods which is based on mapping time series to the graph (a complex network) [10-30]. Two advanced research areas are combined under this approach: the methods of the nonlinear time series analysis [1-8] and the theory of the complex networks [31-38]. There is an opportunity to apply the rich, well developed methods of complex networks analysis to the analysis of the time series with a complex structure, such as the fractal time series.

Currently there are several algorithms for mapping the time series to complex networks. It was suggested in [11] to build a network using the proximity of the coordinates in the Poincare section of the time series (see also [10,12-15]); the algorithm for constructing a natural visibility graph (NVG) was proposed in [16]. Somewhat later, the horizontal visibility graph (HVG) algorithm [17] was proposed. The HVG algorithm is similar to the NVG algorithm but with the slightly modified mapping criterion.

The use of the NVG and the HVG algorithms allow to describe and explore the time series of complex structure associated with a variety of phenomena: fluctuations of turbulent flows [18], stock market indices [19], human heartbeat dynamics [20,21], stochastic and chaotic series [22 - 25] and others [26-30]. In NVG and HVG algorithms each time series corresponds to a single graph.

In [9] a generalization of NVG-algorithm is proposed - the algorithm to construct the graph of dynamical visibility (Dynamical Visibility Graph, DV graph, DVG-algorithm). Each link of NVG is assigned a quantity called the "angle of view". DV graph uses NVG links with "angle of view" less than the specified value $\alpha$, so that every "angle of view" $\alpha$ generates DV graph. In such a way we are able to investigate the properties of the DV graphs upon the angle of view $\alpha$ (rate of growth, jumps etc.). Actually, the ability to change arbitrarily the angle of view $\alpha$ adds the word "dynamic" to the name of the algorithm. We will also use the abbreviation DVG($\alpha$).

Initially, the mapping criterion for the dynamical visibility graph algorithm is described in detail, then we present a set of time series (artificial and experimental) that will be investigated using DVG algorithm. We demonstrate that relative number of clusters in DVG ($\alpha$) graphs at the angle of view $\alpha \to \alpha_c = \pi/2$ behaves similarly to the order parameter in the second-order phase transitions theory. Values of corresponding critical indices are numerically calculated.

We start the construction of the complex network from a sequential time series $\{t_i, i=1..N\}$ in which some events, such as heart beats, occur, i.e. if $i < j \leq N$, then $t_i < t_j$. First we will convert the above time series to $\{x_i = t_{i+1} - t_i, i=1..N-1\}$, therefore $x_i$ values are always positive. In this case we understand that an event "of height" $x_i$ occurs at time $t_i$.

Let us look to the fragment of the time series $t_i, t_{i+8}$ DVG mapping illustration presented on Fig.1. According to [16] the NVG is created from the mapping of a time series of $N$ data to a network of the $N$ nodes and constructed as follows. The points $t_i$ are marked on the horizontal axis and from them in perpendicular direction the line segments with a length $x_i$ are set up. The outer end points of the line segments can be treated as the nodes of the NVG and DVG. A link between two nodes belongs to the NVG if the line connecting the end points of line segments (the NVG nodes) intersects no segments between them. In such a way of NVG construction one can find an angle between connecting lines and vertical axis – "angle of view" of the NVG link. DV graph links are only those with "angle of view" less than the specified limit $\alpha$,

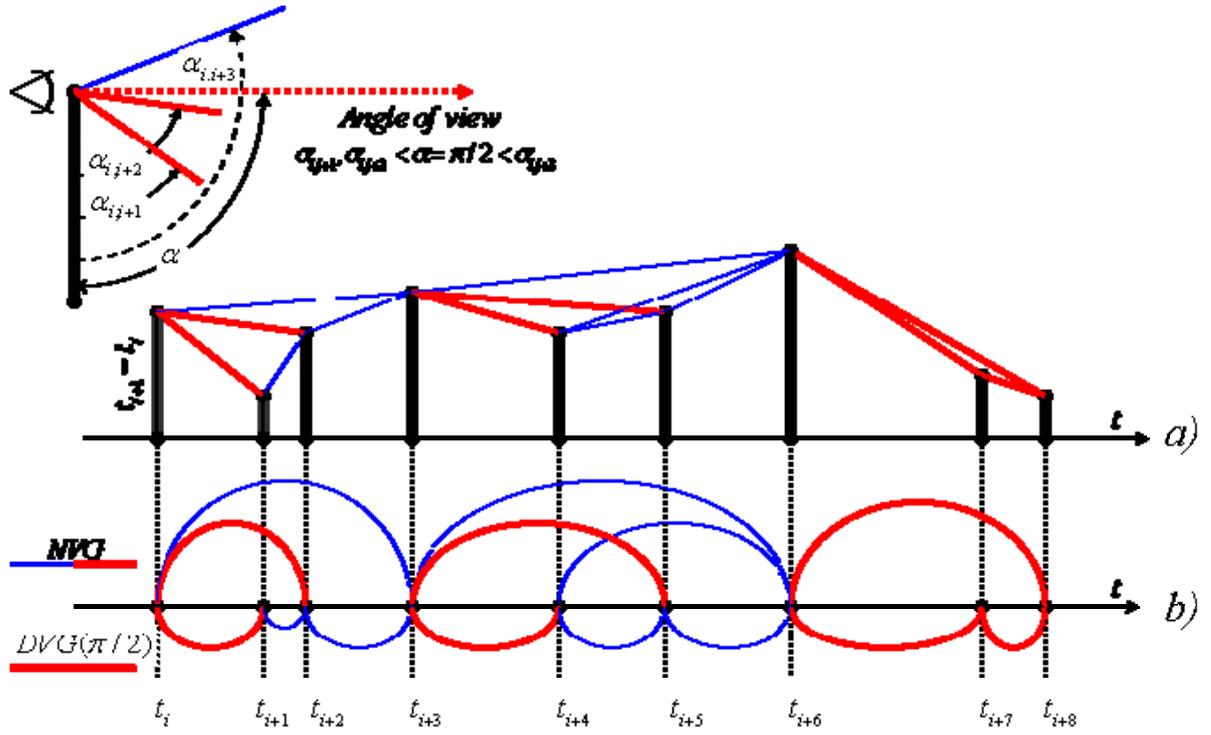

*Fig.1. Illustration of the visibility graph algorithms.* The horizontal axis is a time axis with the marked sequential time events $t_i..t_{i+8}$. Upper left – the DVG link selection criterion for the angle of view $\alpha = \pi/2$ applied to i-th node (a). Lines $t_i, t_{i+1}$ and $t_i, t_{i+2}$ with $\alpha_{i,i+1}, \alpha_{i,i+2} < \pi/2$ (thick lines, red online), line $t_i, t_{i+3}$ with $\alpha_{i,i+3} > \pi/2$ (thin line, blue online). NVG includes all links - both thin and thick lines, DVG includes only link marked by thick lines with angle of view $\alpha_{i,j} < \pi/2$ (b)

Next we specify a formal criterion of visibility for the DVG($\alpha$) i.e. the condition under which the link of the NVG will belong to the DVG($\alpha$). We will consider two arbitrary time events $t_i$ and $t_k$, $i < k$ and all events between them $t_j, i < j < k$. The mapping criterion of the DVG($\alpha$) is a modified visibility criterion [16] of the NVG (1), supplemented by the condition limiting the angle of view $\alpha$ (2):

$$x_k < x_i + (x_j - x_i)\frac{t_k - t_i}{t_j - t_i}, \quad i < j < k, \tag{1}$$

$$\alpha > \alpha_{ik} = arctg\frac{x_k - x_i}{t_k - t_i} \tag{2}$$

We shall consider the following property of DVG-graph - the relative number of clusters in the graph $Q(\alpha)$. Under the cluster we mean here the union of two or more nodes linked acc. to DVG($\alpha$) rules. Single node is considered as a cluster also. Thus, $Q(\alpha)$ - the number of clusters in DVG ($\alpha$) graph divided by the total number of nodes in the whole graph.

It is clearly seen from Fig.1, that at a small angle of view $\alpha$ (look down) no links exist between nodes and the number of clusters is equal to the number of nodes i.e. the relative number of clusters is $Q(\alpha < \pi/4) = 1$. As the angle of view getting wider than $\alpha = \pi/4$ and DVG($\alpha$) graph takes in links from NVG, the average cluster size increases and number of clusters decreases monotonically. At critical angle of view $\alpha_c = \pi/2$ almost all of the nodes are connected, the DVG($\alpha = \pi/2$) graph consists of a small number of clusters and $Q(\alpha \geq \alpha_c) \approx 1/N$ and $N \gg 1$ that means in practice $Q(\alpha \geq \alpha_c) \simeq 0$.

To investigate $Q(\alpha)$ property of DVG($\alpha$) graphs we selected three different random distributions and three experimentally observed time series data. For the construction of artificial time series we used random uniform distribution (3a), the Poisson distribution (3b), and absolute value of Weierstrass (3c) distribution, having a fractal dimension.

$$t_i = t_{i-1} + r, \qquad x_i = r, \qquad (3\ a)$$

$$t_i = t_{i-1} - 1/\lambda \ln(r), \qquad x_i = -1/\lambda \ln(r), \qquad (3\ b)$$

$$a_i = \left| \sqrt{2}\sigma \frac{\sqrt{1-b^{2D-4}}}{\sqrt{1-b^{(2D-4)(M+1)}}} \sum_{m=0}^{M} \left[ b^{(D-2)m} \sin\left(2\pi(sb^m i + r)\right) \right] \right|,$$
$$t_i = t_{i-1} + a_i, \qquad x_i = a_i, \qquad (3\ c)$$

where $r$ - a random variable distributed uniformly on the interval $[0,1]$, following numerical values were chosen: the parameter of the Poisson distribution $\lambda = 1$, the parameters of Weierstrass distribution $\sigma = 3.3$, $b = 2.5$, $s = 0.005$, $M = 10$, $D$ - the fractal dimension. It should be noted that we use the modulus of the values of

Weierstrass distribution because of DVG-algorithm is applicable only to positive values. Further we will call (3c) as WeiMod distribution

Figure 2 shows the time series for the random (3a) and a schematic behavior of the order parameter.

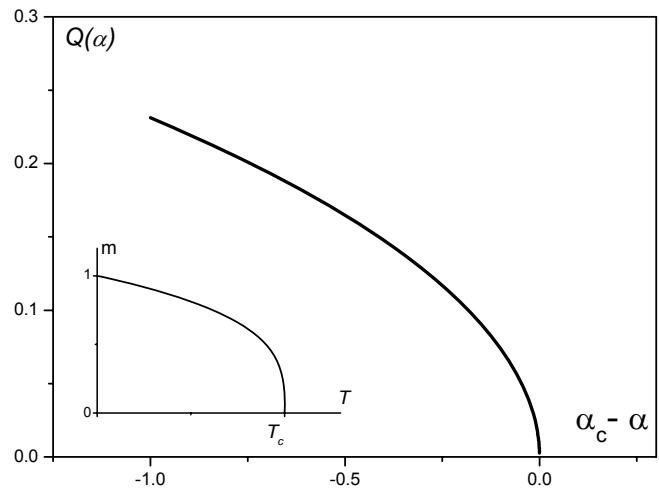

**Fig.2. Plot of** $Q(\alpha)$ **vs.** $\alpha_c - \alpha$ **for a artificial time series (3a).** *In the inset to the figure, the similar plot of the order parameter in second-order phase transition process is presented . The order parameter* $m(T)$ *- the specific magnetization (arbitrary units) as a function of temperature -*$T$ *is plotted,* $T_c$ *- critical temperature (Curie temperature).*

For comparison, the inset of Figure 2 shows a similar dependence for the order parameter in second-order phase transition. In the case of ferromagnetic-paramagnetic phase transition the order parameter $m(T)$ is the specific magnetization, which vanishes as we approach the critical temperature $T_c$ [39-43]. Note that there is an similarity with theory of percolation. In the case of percolation [44,45], the order parameter is the density of the infinite cluster, the probability that a randomly selected node belongs to an infinite cluster and the role of temperature plays the concentration of sites or bonds.

The behavior of the order parameter near the critical value [39-43] is a power low (4), and the exponent $\beta$ is critical index, the main characteristic of the phase transition process.

$$m(T) \sim (T_c - T)^{\beta}. \tag{4}$$

According to Landau [39] mean field theory $\beta = 1/2$, considering fluctuations of the order parameter [42] near the transition point $\beta \approx 0.3$.

In our case, the numerical simulation shows that the relative number of clusters $Q(\alpha)$ behaves near the critical value similarly to order parameter, i.e. obeys power law (see Fig. 2).

$$Q(\alpha) \sim (\alpha_c - \alpha)^{\beta}. \tag{5}$$

Therefore the value of the critical exponent $\beta$ is a characteristic of the time series. Different time series have different $\beta$ values.

We process DVG($\alpha$) for artificial time series (3a-3c) with length $N = 10^5$. Fig. 3 shows the results obtained by averaging over 10 realizations for series (3a-3c). Investigated experimental time series include following. Series of human normal RR-heart beat intervals, data were taken from the Internet resource physionet.org [46]. The number of samples is $6 \div 11 \times 10^4$ per each RR-intervals series. Result was obtained by averaging 72 RR-sequences. Time series of solar flares observed in 2002-2012 (based on hessi.ssl.berkeley.edu [47]). There were registered more than 60,000 solar flares, start times of each flare produced time series. Dates and times of more than 26,000 earthquakes with a magnitude greater than 3.5 on the Richter scale was used to compose earthquake time series. Data was taken from seismorus.ru [48], events was occurred in 1960 - 2010 In Russia and the CIS region.

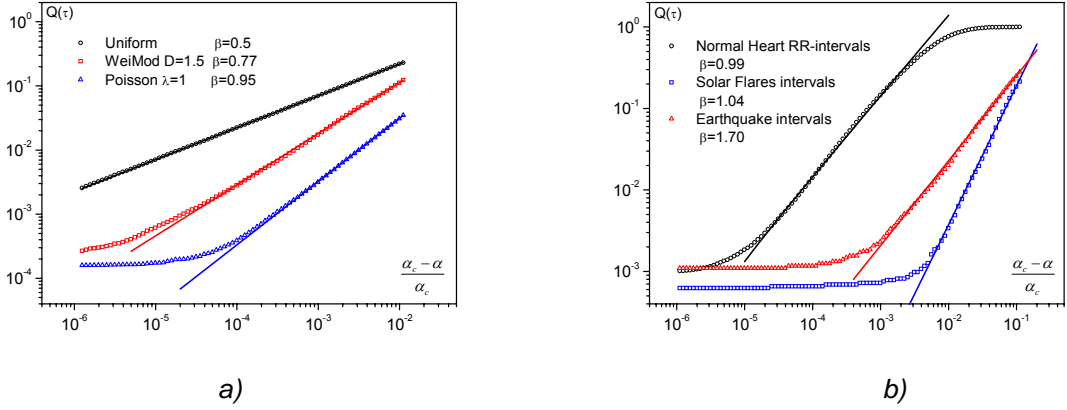

***Fig.3. Relative cluster quantity*** $Q(\alpha)$ ***on the proximity to the critical angle*** $\alpha_c - \alpha / \alpha_c$ ***in a log-log scale. a) artificial series, b) experimental series. Type of time series and value of the critical index indicated in figure.***

According to the results for $Q(\alpha)$ shown in Figure 3, the lowest value of the critical exponent, which coincides with the numerical value of the critical exponent $\beta = 0.5$ of the order parameter in the mean-field theory of Landau [39], has the time series with a random uniform distribution (3a). Critical exponents of other artificial series (3b) and (3c) were greater. Also Fig. 3a shows that $Q(\alpha)$ becomes independent upon the critical angle $\alpha$ near to $\alpha_c - \alpha$. Such behavior of the order parameter near to the phase transition point is well known in second-order phase transitions in the presence of the so-called external field, which smears the phase transition [39-43].

We also calculated $Q(\alpha)$ for series of experimental data: (see Fig. 3b) a number of RR-intervals of the normal human heart rate, series built on intervals between solar flares and between earthquakes. The experimental data processed using DVG algorithm showed that the dependence $Q(\alpha)$ of these series has all of the features of the phase transition like artificial has. It is interesting to note that in all investigated cases the critical exponent $\beta$ was greater than for the artificial distributions. Entering the "plateau" near the transition point begins for experimental distributions by an order earlier than in the artificial. The authors believe that the measurement resolution results in this phenomena. The data for solar flares and earthquakes are shown with a second accuracy and the measurement of heart rate accuracy is 0.01 seconds.

Of course, as it follows from the above comparison of presented results, the time series of different nature (random, with a complex structure, experimental etc.) are characterized by their numerical value of the index $\beta$. The index $\beta$ itself is not enough to determine the type of distribution of the intervals of the time series or the distribution of input data. But the $\beta$ value is next, new parameter (along with the fractal dimension, the Hurst exponent or index of the power spectrum etc.) to describe time series with a complex structure, including both random and deterministic components.

Proposed and shown in this paper similarity of the behavior of the relative clusters quantity in the dynamical visibility graphs and the order parameter in the theory of second-order phase transitions seems a new step towards the investigation of the properties of time series. This similarity will allow in the future to apply to the investigation of time-series the methods developed in second order phase transition theory, to determine the role of the critical behavior of the various parameters of the dynamical visibility graphs for the investigations of time series of different structures.